\documentclass{article}

\usepackage{arxiv}
\usepackage{amsmath}
\usepackage[utf8]{inputenc} % allow utf-8 input
\usepackage[T1]{fontenc}    % use 8-bit T1 fonts
\usepackage{hyperref}       % hyperlinks
\usepackage{url}            % simple URL typesetting
\usepackage{booktabs}       % professional-quality tables
\usepackage{amsfonts}       % blackboard math symbols
\usepackage{nicefrac}       % compact symbols for 1/2, etc.
\usepackage{microtype}      % microtypography
\usepackage{lipsum}
\usepackage{graphicx}
\graphicspath{ {./images/} }

\title{Reconciling Experimental and Theoretical Vibrational Deactivation in Low-Energy O+N$_2$ Collisions}

\author{
 Qizhen Hong \\
  State Key Laboratory of High Temperature Gas Dynamics, \\ 
  Institute of Mechanics, Chinese Academy of Sciences, 100190 Beijing, China and\\
   School of Engineering Science, \\
  University of Chinese Academy of Sciences, Beijing 100049, China
 \\
   \And
 Massimiliano Bartolomei \\
  Instituto de F\'isica Fundamental - CSIC, C/ Serrano 123, Madrid, Spain \\
  \And
 Fabrizio Esposito \\
  Consiglio Nazionale delle Ricerche, Istituto per la Scienza e Tecnologia dei Plasmi,\\
  Sede Secondaria di Bari, 
via Amendola 122/D 70126 Bari, Italy\\
  \And
 Cecilia Coletti \\
Dipartimento di Farmacia, Universit\`a G. d'Annunzio Chieti-Pescara, via dei Vestini, 66100 Chieti, Italy \\
  \texttt{ccoletti@unich.it} \\
    \And
 Quanhua Sun \\
  State Key Laboratory of High Temperature Gas Dynamics, \\ 
  Institute of Mechanics, Chinese Academy of Sciences, 100190 Beijing, China and\\
   School of Engineering Science, \\
  University of Chinese Academy of Sciences, Beijing 100049, China
 \\
     \And
 Fernando Pirani \\
 Dipartimento di Chimica, Biologia e Biotecnologie, \\
 Universit\`a di Perugia, via Elce di Sotto 8, 06183 Perugia, Italy
}

\begin{document}
\maketitle
\begin{abstract}
Molecular dynamics calculations of inelastic collisions of atomic oxygen with molecular nitrogen are known to show orders of magnitude discrepancies with experimental results in the range from room temperature to many thousands of degrees Kelvin. In this work, we have achieved an unprecedented \emph{quantitative} agreement with experiments even at low temperature, by including a non-adiabatic treatment involving vibronic states on newly developed potential energy surfaces. This result paves the way to the calculation of accurate and detailed databases of vibrational energy exchange rates for this collisional system. This is bound to have an impact on air plasma simulations in a wide range of conditions and on the development of Very Low Earth Orbit (VLEO) satellites, operating in the low thermosphere, objects of great technological interest due to their potential at a competitive cost. 
\end{abstract}

% keywords can be removed
%\keywords{First keyword \and Second keyword \and More}

\section{Introduction}
Reactive, inelastic and dissociation processes in molecular collisions of air species play a crucial role for the accurate modelization of air plasmas, which include combustion processes \cite{adamovich_challenges_2015}, planetary entry problems \cite{celiberto_atomic_2016}, electrical discharges \cite{pintassilgo_power_2016}, atmospheric kinetics \cite{pavlov_vibrationally_2011,yankovsky_model_2020}, plasma medicine \cite{keidar_plasma_2015}. In these systems it is common to find strong non-equilibrium conditions in different molecular degrees of freedom, which impose the adoption of detailed state-to-state models \cite{capitelli_thermodynamics_2012} for the comprehension and control of phenomena of wide technological interest. In these models the required input data may come from experiments and/or from theoretical calculations. In the first case it is quite unlikely to extract all the needed data, considering the large total energy ranges normally required. As a consequence, the detailed computations of (ro)vibrationally detailed kinetic data by molecular dynamics methods from accurate potential energy surfaces (PES), describing the interactions in the collisions, become utterly necessary. The results of these calculations can then be compared with the experimental data in the usually available intervals, in order to assess their accuracy. 
Among air species processes, one case of special interest is represented by the collisions of atomic oxygen with molecular nitrogen. The quite common presence of atomic oxygen in air plasmas is due to the fact that molecular oxygen dissociation threshold is much lower than for nitrogen. Atomic oxygen is also found in the low thermosphere, a region between 90 and 250 Km of altitude, of great relevance for future satellite constellations, where UV radiation from the Sun dissociates O$_2$, so that the most abundant species are precisely O and N$_2$. The accurate knowledge of their interactions, determining the low temperature range 300-1000 K behavior, is urgent, in view of the development of VLEO satellites with air-breathing electric propulsion  system \cite{parodi_study_2019,erofeev_air-breathing_2017,leomanni_propulsion_2017,holste_ion_2020}, capable in principle of endless operation without a propellant tank. 
As a consequence, the detailed study of O+N$_2$ interactions is currently of strategic importance. Collisions of O with N$_2$($v$) ($v$/$v^{'}$= initial/final vibrational quantum numbers) in turn generate  excited N$_2$($v^{'}$) molecules in an inelastic process, or NO($v^{'}$)+N reaction, or N$_2$ dissociation. Recent computations of reaction and dissociation rates appear to be in good agreement with what is known about these processes \cite{esposito2,truhlar,luo_ab_2017,meuwly2020a}, 
but  experimental results for the O+N$_2$(1)$\rightarrow$ O+N$_2$(0) inelastic process in the 300-4500 K interval are not yet reproduced by theoretical calculations, which present severe under-estimations (of orders of magnitude) at room temperature \cite{esposito2,meuwly2020a,meuwly2020b}, where the system is only characterized by non-reactive collisions and accurate values are required for aerospace technological applications. Results at temperature higher than 1000 K also present important over-\cite{meuwly2020a,meuwly2020b} or under-estimations \cite{esposito2} .

\section{Potential energy surface}
The basic ingredient needed to properly investigate elementary processes promoted by collisions is the accurate characterization of the interaction potentials driving the molecular dynamics. The evolution of inelastic scattering processes, particularly at low temperature, is known to be strongly dependent on the long range region of the potential, a part of the PES which is seldom well characterized.  Therefore, the present investigation was initially motivated by the need of an accurate non-reactive PES, providing the best possible description at long and medium range, and likely to give a physically meaningful insight on inelastic collisional events. 
 Such potential in turn should be expressed in a simple form capable to represent the full space of relative configurations of the involved partners. It should furthermore describe the formation, by two-body collisions, of weakly bound adducts, representing the precursor states of further basic processes. Note that this might still be a tough challenge for potentials exclusively based on ab-initio computations: a very high level of theory is required to evaluate the small interaction energies of weakly non-covalently bound systems and the number of points required to fully cover all possible long range regions could easily become prohibitive.

We have thus represented the multidimensional PES for O($^3$P$_J$)-N$_2$($^1\Sigma_g^+$) in analytical form,
by taking into account that at intermediate and large intermolecular distances $R$ the interaction is determined
by the balance between van der Waals (vdW) forces and other contributions,  deriving from different reciprocal alignments of N$_2$ molecular axis and
 of the half-filled orbitals of O($^3$P$_J$) atom with respect to $R$ \cite{Ref1}. $J=2,1,0$ represent the total electronic angular momentum states of the oxygen atom, which, for a plasma at T$\ge$ 1000 K, are statistically populated in 5:3:1 ratio. In the following we will use the shorter notation  O($^3$P) for simplicity.  

 Strength and anisotropy of the interaction contributions other than $V_{vdW}$ are mainly dependent on the electrostatic quadrupole-quadrupole component (V$_{el}$), arising from the non-spherical electronic charge distribution of both partners, and on selective charge transfer (CT) effects (V$_{ct}$)
 in the perturbation limit emerging in systems involving high electron affinity open shell atoms, as is the case for O($^3$P) \cite{Ref1,Ref2}. 
Accordingly, we have defined the total interaction V$_{tot}$ as 

\begin{equation}
  V_{tot}=V_{vdW}+V_{ct}+V_{el}
\end{equation}

Particular effort has been addressed to represent each of the contributions through simple analytical formulae depending on few and physically meaningful parameters, leading to a correct representation of the interaction in the full space of the relative configurations.

 V$_{el}$ is given by the canonical expression of quadrupole-quadrupole interaction, whereas the sum of the first two components, $V_{vdW}+V_{ct}$, has been formulated as the combination of pair interactions between O($^3$P) and each N atom of the N$_2$ molecule, described by an Improved Lennard Jones (ILJ) function \cite{Ref3}, whose details are given in the Supplemental Material (SM).
Zero order values of the parameters involved in the ILJ expression have been estimated from the polarizability of the O 
atom (0.8 \AA$^3$) and the effective component of that of each N atom  (0.9 \AA$^3$)  within N$_2$. 
According to the ample phenomenology of O($^3$P) interacting with closed shell partners \cite{Ref2,Ref4}, 
two different types of interaction can be distinguished: the 
  oxygen atom approaching N$_2$ with one of the half-filled $p$ orbitals aligned parallel to the intermolecular distance
 $R$, leading to the formation of $^3\Pi$ state (electronic molecular quantum number $\Lambda=1$), and that with the oxygen atom approaching with the only filled orbital aligned along $R$, leading to a  $^3\Sigma$ state ($\Lambda=0$)  \cite{Ref2}.
This diversification accounts for the contribution of CT, which was found to selectively stabilize states of $\Pi$ symmetry \cite{Ref4}.

The zero order parameters have then  been fine tuned exploiting  the simultaneous comparison of the predicted intermolecular interaction 
with the results of ab-initio calculations, and its ability to reproduce experimental total cross sections, leading to the values reported in Table S1 in SM.

\begin{figure}
\centering
\includegraphics[width=10cm,angle=0.]{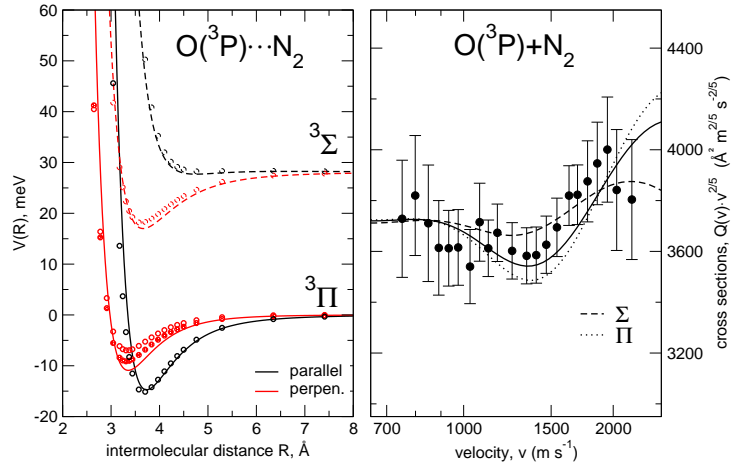}
\caption[]{ Left panel: Intermolecular potentials for the  interaction between O($^3$P) and N$_2$($^1\Sigma_g^+$) as a function of the distance $R$ between the oxygen atom and the center of mass of the N$_2$ molecule at its equilibrium distance (r$_e$=1.1007 \AA). Black and red lines are obtained through the present analytical PES and
 correspond to the parallel and perpendicular orientations, respectively, of N$_2$ with respect to the oxygen atom. 
Symbols, colored accordingly, correspond to the ab-initio calculations carried out at the  CCSD(T)/CBS level of theory. 
Right panel: Integral cross sections for the O($^3$P)+N$_2$($^1\Sigma_g^+$) collisions  as a function of the selected atom beam velocity $v$. Full circles correspond to experimental data from Perugia laboratory \cite{Ref5}, while curves to calculations onto the present analytical PESs using an Infinite Order Sudden approximation.
}
\label{fig1}
\end{figure}

The left panel of Fig.\ref{fig1} reports the two O($^3P$)-N$_2$($^1\Sigma_g^+$) ground $^3\Pi$  and first excited $^3\Sigma$ (lying asymptotically 28.1 meV above the $^3\Pi$ state and directly correlating with the excited spin orbit level $^3$P$_0$ \cite{Ref2,Ref4}) PESs as a function of the intermolecular distance $R$ for the two limiting orientations (parallel and perpendicular) of the diatom (at its equilibrium distance)  approaching the oxygen atom.

The figure also shows ab-initio energy values obtained at the CCSD(T)/CBS level of theory. Note that, in the case of the perpendicular orientation for the $^3\Pi$ state, ab-initio data provide two non-coincident sets (red circles) for the potential energies, depending on the two different orientations of the fully occupied $p$ orbital of the oxygen atom with respect to the diatom. 
The two sets are degenerate under the C$_{\infty v}$ (parallel) configuration of the system and split in the C$_{2v}$ (perpendicular) configuration, corresponding to the $^3$B$_2$ and $^3$B$_1$ symmetries. Such symmetries in turn correlate with the $^3$A$^{'}$ and  $^3$A$^{''}$ states of the more general C$_s$ configuration.
Because the difference in energy between these two kinds of interaction is small and tends to zero for both small and long $R$ values (and does not exist in the $^3\Sigma$  state), we simplified the present model by only considering an average contribution of the $^3$A$^{'}$ and  $^3$A$^{''}$ states, collectively indicating it as $^3\Pi$ PES.

Fig. \ref{fig1} (right panel)  reports the experimental total cross sections $Q$, measured as a function of the selected velocity $v$ of projectile O atoms and under single collision conditions with the target N$_2$ molecules \cite{Ref5}. The data have been plotted as $Q(v)\cdot v^{2/5}$ to emphasize the quantum interference effects, observable as an oscillatory pattern in the $v$ dependence of  the measured $Q(v)$.
The  average $Q(v)$ values directly probe  the strength of 
the long range average dispersion attraction, while the extrema position and the frequency of the oscillating pattern give unique information on the depth and on the minimum location of the potential well, which occur at intermediate separation distances and are determined by 
a critical balance of attraction and repulsion. 
The total cross sections calculated on both $^3\Pi$ and $^3\Sigma$ PESs are also displayed in the figure, together with their combination, weighted 
by the degeneracy ratio 2:1, according to the statistical population of the oxygen fine levels in the
 atomic beam \cite{Ref5} (see SM). The latter is shown to reproduce the relevant behavior of the experimental data within their uncertainty.

\section{Vibrational Relaxation in N$_2$ + O collisions}
\begin{figure}
 \centering
\includegraphics[width=10.cm]{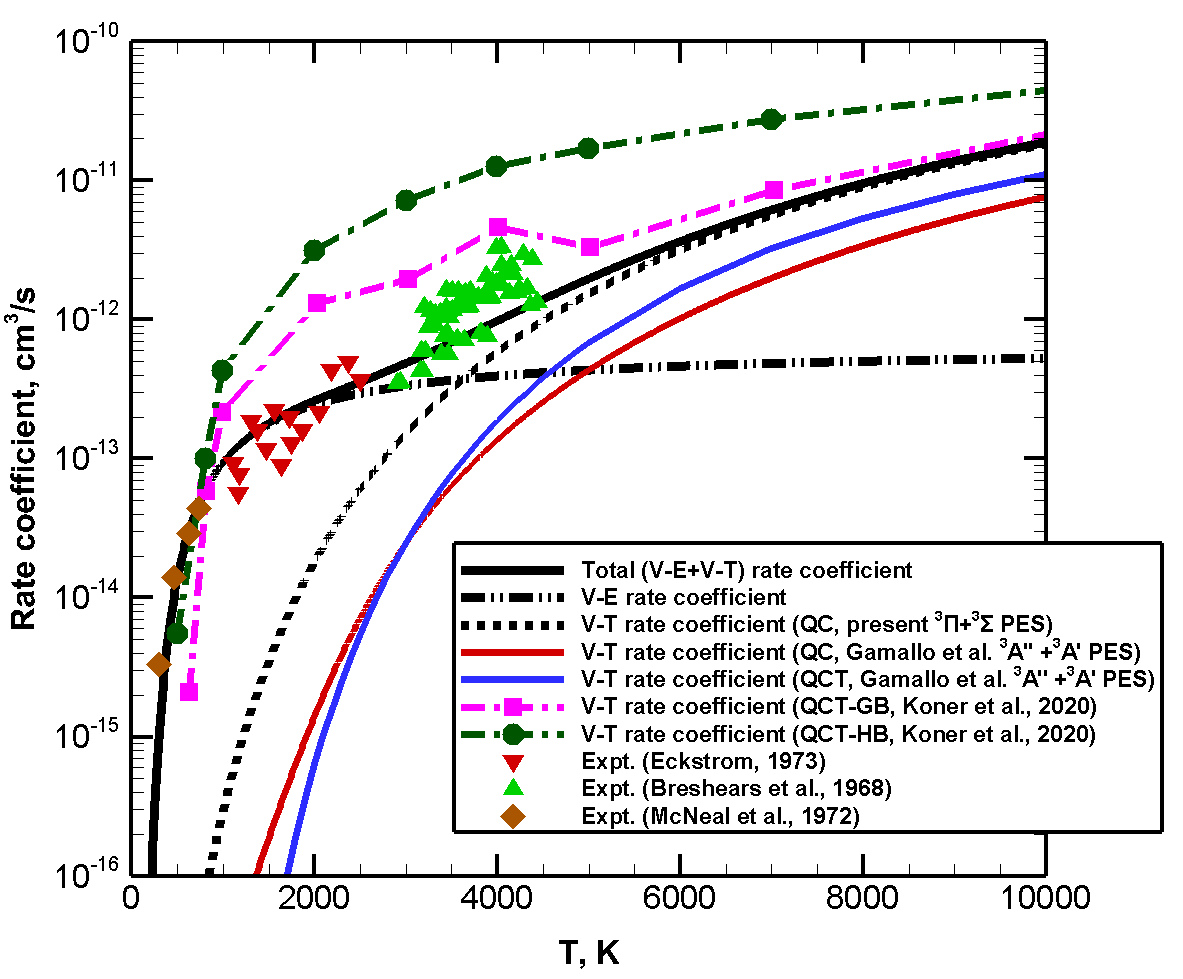}
\caption[]{Rate coefficients for vibrational relaxation upon O($^3$P)-N$_2$($^1\Sigma_g^+$)($v=1$) collision as a function of temperature. Experimental data  by Eckstrom \cite{eckstrom} (red down triangles), by Breshears and Bird \cite{breshears} (green up triangles) and by McNeal et al.\cite{mcneal} (brown diamonds) are reported together with QCT\cite{esposito2} (blue solid line)  and QC (red solid line) V-T rate coefficients computed on Gamallo et al. PES \cite{sayos}, on Koner et al. PES \cite{meuwly2020a,meuwly2020b} by a QCT method associated with a standard Histogram Binning (dark green dash-dot line with circles) or a Gaussian Binning procedure (pink dash-dot line with squares), and on the present $^3\Pi$ and $^3\Sigma$ surfaces (2:1 averaged) with a QC method (black dotted line). Rate coefficients for the non-adiabatic vibro-electronic (V-E) transition between $^3\Pi (v=1)$ and $^3\Sigma (v=0)$ calculated according to the LZ approach are drawn as a black dashed-dotted line. The weighted sum of the present V-T and V-E rate coefficients, representing the total vibrational relaxation rate, is reported as a solid black line.    
}
\label{fig3}
\end{figure}

The present non-reactive PESs, with the inclusion of the intramolecular potential for N$_2$ (here taken as the Morse potential)  can be used for the calculation of the rates of inelastic vibrational relaxation processes involving vibrationally excited N$_2$ molecules at temperature lower than 10000 K, where the influence of the reactive channels is still small. All the computational details are given in the SM.
 The case of the vibrational relaxation of N$_2$($v$=1) upon collision with atomic oxygen is particularly intriguing because quasi-classical trajectory (QCT) calculations performed on most of the existing PESs \cite{sayos,truhlar,meuwly2017}, available for the $^3$A$^{'}$ and  $^3$A$^{''}$ states of this system, are known to underestimate \cite{ivanov,esposito2} the experimental rate coefficients of 1-2 orders of magnitude at T$\approx$ 2000-4000 K and of 3-4 orders of magnitude at lower temperature (see Fig. \ref{fig3}). At very low temperature (less than 500-700 K) QCT calculations are not able to foresee any probability of V-T energy exchange. This is expected, because vibrational inelastic energy transfer can be a {\em classically forbidden} process in the sense of Ref. \cite{miller_classical_1970}.  In short, the classical final vibration is only slightly different from the initial one for sufficiently low energy so that the QCT binning becomes unable to detect a small, but non-zero, result different from the elastic one \cite{esposito2}.
 In order to avoid this effect, which could affect the results independently on the PES quality, we used a mixed quantum-classical (QC) method \cite{billing1984rate,noi1,noi2}, whereby the N$_2$ vibration is described by quantum mechanics and the other degrees of freedom classically. We used the QC method also in combination with Gamallo et al. PES \cite{sayos}, for which QCT calculations are available \cite{esposito2}, allowing for the quantification of the effect of the QC dynamical treatment over the QCT one, as also reported in Figs. \ref{fig3} and S1 in SM. As expected, QC and QCT values are close at T$\ge$ 3000 K, whereas QC results grow larger than QCT ones as temperature decreases. 
 %Unfortunately, QCT issues alone are clearly not able to justify the low-temperature discrepancy with experimental data.
 
Fig. \ref{fig3} also reports the rate coefficients calculated on the present ground $^3\Pi$ and excited  $^3\Sigma$ PESs averaged according to their statistical 2:1  population. Rate coefficients calculated separately on the $^3\Pi$ and on the $^3\Sigma$ potentials can be found in Fig. S2 in SM. Only those obtained on the $^3\Pi$ PES can be directly compared to those calculated on the $^3$A$^{'}$ and $^3$A$^{''}$ PESs.
The new PESs, and in particular the $^3\Pi$ one, provide sensibly higher (ca. one order of magnitude at T$\ge$ 2000 K and larger as the temperature lowers) rate coefficients than Gamallo et al. PES \cite{sayos}, whereas the standard QCT values calculated with the recent Koner et al. PES \cite{meuwly2020a,meuwly2020b} are up to three orders of magnitude larger in the 1000-4000 K interval. If Gaussian binning is instead used, their QCT rates are closer at high temperature, but the discrepancy becomes significantly worse at low temperature.
Compared to the experimental vibrational relaxation data, the present results are only slightly smaller in the range 3000-4000 K, but the difference, strongly growing as temperature decreases, rises up to 2-3 orders of magnitude at T$\le$ 3000 K. 
This behavior and a closer look at the $^3\Pi$ and $^3\Sigma$ PESs point out that the reason for such apparent discrepancy must have a different origin. 

\begin{figure}
\centering
\includegraphics[width=10.cm]{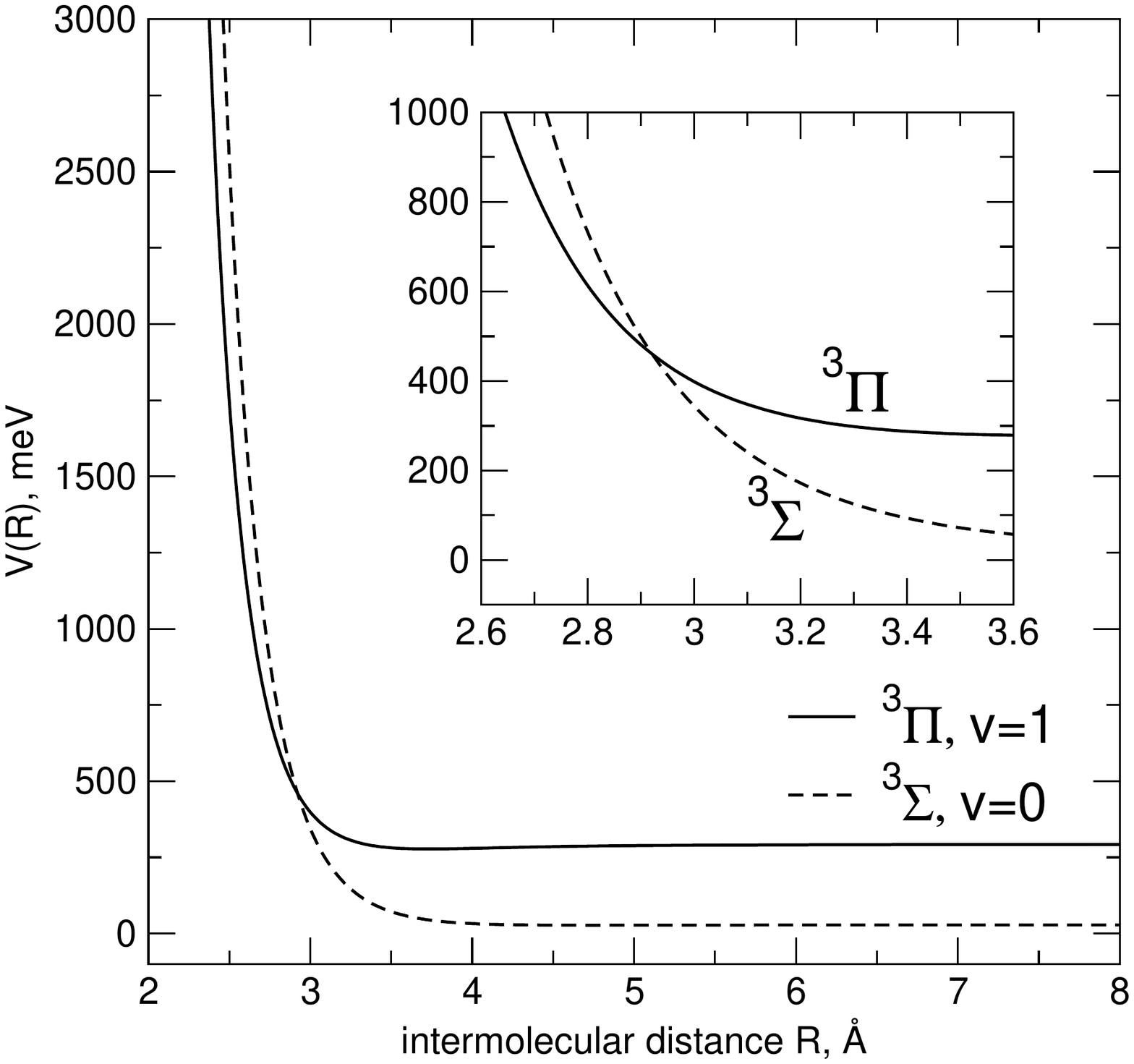}
\caption[]{ Potential curves for the O($^3$P)-N$_2$($^1\Sigma_g^+$) interaction with N$_2$ molecular axis oriented along the intermolecular distance. In order to determine the main features between vibronic states responsible for molecular relaxation,  the molecule is assumed in the first excited  $v$=1 vibration level for the $^3\Pi$ state, while for the $^3\Sigma$ state $v$=0 is considered.  
}
\label{fig2}
\end{figure}

As suggested by Nikitin and Umanski \cite{nikitin1}, the unusually high vibrational relaxation rates for N$_2$($v$=1) when colliding with O($^3$P) at low T is due to the open shell nature of oxygen leading to the non-adiabatic vibro-electronic (V-E) energy transfer \cite{candori1,candori2} which takes place at the crossing between the $^3\Pi$ and $^3\Sigma$ vibronic surfaces. Indeed, as shown in Fig. \ref{fig2} for the collinear C$_{\infty v}$ configuration, expected to be the most effective for inelastic events promoted by vibronic couplings, the $^3\Pi$($v$=1) and the $^3\Sigma$($v$=0) PESs cross at $R_c$=2.922 \AA\ where the potential energy is $E_x$=0.1666 eV in the entrance channel, a value obtained as the difference between the energy at the crossing $E_c$, 0.4592 eV, and the vibrational quantum of energy for N$_2$($v$=1), 0.2926 eV.

The detailed description of the two PESs gives us the possibility to quantitatively evaluate the V-E contribution to the vibrational quenching rate, according to the Landau-Zener (LZ) approach \cite{landau,zener,stuckelberg}.
V-E rate coefficients are reported as a function of temperature in Fig. \ref{fig3}: they strongly increase  with temperature to reach a plateau at T$\ge$2000 K. At temperature lower than 2000 K, V-E rates are higher than V-T ones, and the overall vibrational quenching rate is thus mainly determined by the vibronic energy transfer process. This is very clearly indicated by the calculated total relaxation rate (Fig. \ref{fig3}), obtained as the sum of the V-T contribution and the V-E one, the latter multiplied by 2/3 as it only occurs on the $^3\Pi$ PES. The excellent agreement between calculations and experimental data (Tab. S3 and Fig. S3 in SM), both at low (where V-E energy transfer dominates) and high (where V-T rate coefficients prevail) temperature, represents a strong indication that the apparent theory-experiment disagreement is in fact the result of the neglect of one important physical contribution to the removal of excited nitrogen molecules. 
 Note that the non-adiabatic transition between the reactants triplet PESs and the N$_2$O singlet \cite{fisher_quenching_1972}, occurring at higher energies (with a threshold around 1 eV), might be responsible for the slight difference between the present relaxation rate and experimental data in the temperature range over 3500 K.

\begin{figure}
\centering
\includegraphics[width=10.cm,angle=0.]{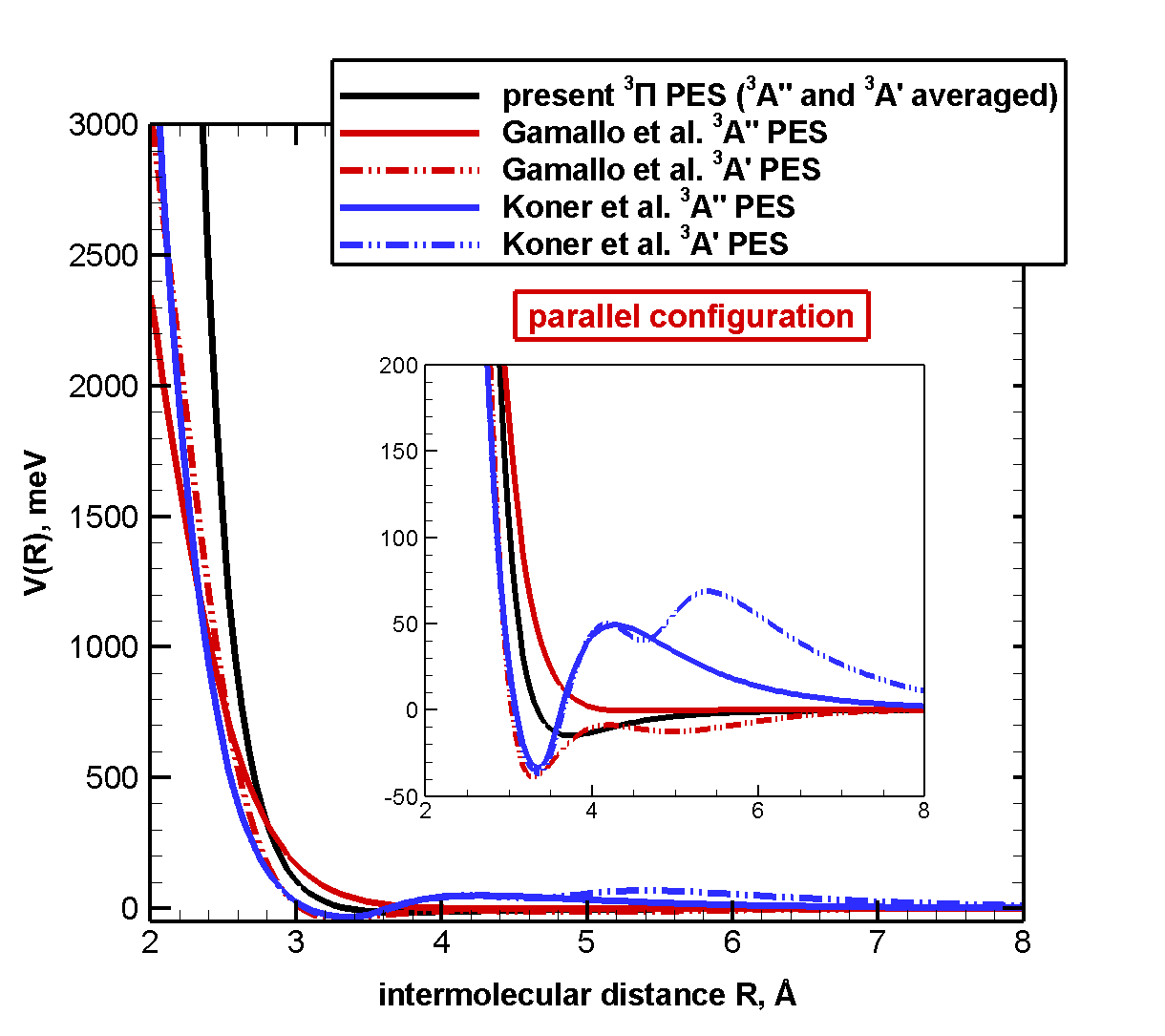}
\caption[]{Behavior of different potential energy surfaces as a function of the intermolecular distance $R$ for the collinear (or parallel) configuration, corresponding to the  C$_{\infty v}$ symmetry. The present $^3\Pi$ PES is reported as a solid black line, the Gamallo et al. \cite{sayos} $^3$A$^{''}$ and $^3$A$^{'}$ are the red solid and dashed lines, respectively, and the Koner et al. \cite{meuwly2020b} $^3$A$^{''}$ and $^3$A$^{'}$ are the blue solid and dashed lines, respectively.
}
\label{fig4}
\end{figure}

The matching between calculated and experimental values of total relaxation rates prompts us to investigate where the discrepancy between V-T rate coefficients computed on the present PES and those available in the literature (Fig. \ref{fig3}) arises from.
To this aim Fig. \ref{fig4} reports the potential energy as a function of the intermolecular distance $R$ for the $^3$A$^{'}$ and $^3$A$^{''}$ PESs of Gamallo et al. \cite{sayos} and those of Koner et al. \cite{meuwly2020b} and the present $^3\Pi$  for the parallel configuration, the most relevant for the processes considered here. We recall that $^3$A$^{'}$ and $^3$A$^{''}$ should be degenerate for the parallel (collinear) configuration and asymptotically for all configurations. 

 Fig. \ref{fig4} 
shows that in fact the $^3$A$^{''}$ and $^3$A$^{'}$ PESs of ref. \cite{sayos} only coincide at long range, the short range divergence probably due to the interpolation procedure which, by mixing the C$_{\infty v}$ collinear points with diverse symmetry non-degenerate neighbouring points  ($^3$A$^{''}$ and $^3$A$^{'}$ differently correlate  with the reactive channels), might spuriously remove the degeneracy. The qualitative behavior of both PESs however is similar to the new $^3\Pi$ surface (falling below and above respectively) at long range, with V-T rates calculated on the $^3$A$^{'}$ slightly larger (Fig. S1 in SM). Both $^3$A$^{''}$ and $^3$A$^{'}$ surfaces at short range are less repulsive than $^3\Pi$ potential which might be the reason why V-T rate coefficients are about one order of magnitude smaller than the presently computed ones.

The $^3$A$^{''}$ and $^3$A$^{'}$ PESs of ref. \cite{meuwly2020a} show the opposite behavior: they practically coincide at short range (up to R $\approx$ 4.5 \AA), but they diverge at long range, where they both present a high early barrier with at least 50 meV peak value and a well with a steep attractive side. The very large value of V-T rate coefficients calculated on these PESs, starting from 1000 K, can  likely be attributed to such repulsive behavior at long interaction distances, which are relevant on the outcome of V-T dynamics calculations. 
%If this is due to special features of accurate ab-initio calculations or to inaccurate fitting has to be determined with further studies. 
Similar differences can be found for the perpendicular configuration as shown in Fig. S4 in SM.

\section{Conclusions}
The present investigation, providing a detailed characterization of intermediate and asymptotic regions of the O($^3$P)-N$_2$ interaction, casts light on the presence of crossings between potential energy surfaces of different electronic symmetry where vibronic non adiabatic events are triggered. Therefore, in addition to the canonical vibration-translation inelasticity, vibration-electronic energy transfers can also be effectively promoted by collisions. One crucial point here is that the two types of events emerge in different ranges of gaseous-mixture temperature. Note that the significance of these findings is not limited to O($^3$P)-N$_2$ collisions, they are of broad interest for the control of elastic and inelastic elementary processes occurring in several plasmas where open shell O atoms are involved in the collision with many other molecular partners. 

%\begin{table}
% \caption{Sample table title}
%  \centering
%  \begin{tabular}{lll}
%    \toprule
%    \multicolumn{2}{c}{Part}                   \\
%    \cmidrule(r){1-2}
%    Name     & Description     & Size ($\mu$m) \\
%    \midrule
 %   Dendrite & Input terminal  & $\sim$100     \\
%    Axon     & Output terminal & $\sim$10      \\
%    Soma     & Cell body       & up to $10^6$  \\
%    \bottomrule
 % \end{tabular}
%  \label{tab:table}
%\end{table}

\emph{Acknowledgments}
Q.H. and Q.S. acknowledge financial support from the Strategic Priority Research Program of Chinese Academy of Sciences (Grant No. XDA17030100) and the National Natural Science Foundation of China through grants 11372325 and 91116013. M.B. acknowledges the FIS2017-84391-C2-2-P Spanish grant for fundings.

\section{Supplemental Material}
\subsection{Experimental and Computational Details}
Ab-initio calculations of the O($^3\Pi$)-N$_2$($^1\Sigma_g^+$) intermolecular interaction energies have been carried out at the CCSD(T) level of theory by using the Molpro code \cite{Molpro} and the computed values have been corrected for the basis set superposition error by the counterpoise method of Boys and Bernardi \cite{Boys:70}. The complete basis set (CBS) extrapolation of the obtained interaction energies has been performed by exploting the two-point correlation energy procedure of Halkier et al.\cite{Halkier:98,Halkier:99} in conjuction with Dunning \cite{Dunning} augmented correlation-consistent aug-cc-pVQZ and aug-cc-pV5Z basis sets. 
For the analytical representation of the $^3\Pi$  and $^3\Sigma$ PESs we have considered that the sum of first two terms in Eq. 1 (V$_{vdW}$+V$_{ct}$) is globally accounted for with the sum of O-N atom-effective atom contributions, each one represented by an ILJ formula \cite{Ref3}, while the last term (V$_{el}$) is described by a canonical expression of the quadrupole-quadrupole interaction, due to the non-negligible quadrupole moments of both monomers, and the related parameters are reported in Table S1.
The internal coordinate dependence of both N$_2$ polarizability and quadrupole moment is taken from Refs.\cite{cappelletti:08,CO2-N2:16}.

Molecular beam scattering experiments have been performed several years ago in the Perugia laboratory
 with an apparatus described in detail elsewhere \cite{Ref5}. In short, high angular and velocity resolution conditions 
 were adopted in order to measure quantum \emph{glory} interference effects, observable as an oscillatory pattern in the velocity dependence 
of the integral cross section $Q(v)$. A microwave discharge source operating at low pressure (few mbar) and temperature of about 10$^3$ K was used to
 generate the oxygen beam formed by atoms in their ground $^3$P$_J$  electronic state with a near statistical population of the
  spin orbit levels J=2,1,0. It has been also shown\cite{Ref4} that while J=2 and J=0 correlate at intermediate and short $R$  with states of $\Pi$ and
 $\Sigma$ character, respectively, J=1 provides a combination of both; this assures that  states of the two different symmetries
 are formed in a 2:1 statistical ratio during the scattering.  The target gas formed by N$_2$ molecules was contained in a scattering chamber cooled at about 90 K. 
 During the present analysis, which overcomes the previous one performed by a simple spherically symmetric potential model \cite{Ref5},  
cross sections are calculated from the present anisotropic intermolecular potential in the center of mass system within the semiclassical 
JWKB method and convoluted in the laboratory frame for a critical comparison with the experimental results. Taking into account 
that the rotational motion of molecules is cooled at about 90 K and that the proposed interaction is strongly anisotropic, the infinite order sudden (IOS) approximation, that considers collisions occurring at fixed relative orientation of partners, is adopted.

For the quantum-classical calculations \cite{billing1984rate} in this work (see the following), the lowest 9 vibrational states of N$_{2}$ are used and the initial state is $v$=1. The rates are computed at 47 initial values of total classical energy comprised between 50 cm$^{-1}$ and 80000 cm$^{-1}$, with a more frequent sampling directed towards lower energies. For each energy value, 5000 trajectories were used, as well as an initial separation distance atom-diatom $R$ equals to 50 \AA\ and an impact parameter randomly chosen between 0 and 9 \AA.

The probability of V-E transfer $P_{x}$ is calculated according to the well-known Landau-Zener procedure \cite{landau,zener,stuckelberg}:

\begin{equation}
\label{prob}
P_{x}=\exp \left(-\frac{2 \pi H^{2}}{\hbar v_{R} \Delta}\right).
\end{equation}
where $v_R$ is the radial velocity at the crossing, $\Delta$ is the difference between the slope of the two PESs at the crossing point, here calculated to be $\Delta=$0.7846 eV/\AA. Note that only the collinear configuration is taken into account, as the most effective for inelastic processes. $H$ is the coupling between the two PESs. In the range of collision energies associated to the temperature of interest here (below 10000 K), the use of a fixed value of $H$ is sufficient to calculate reliable cross sections and rate coefficients. The exact determination of the value of $H$ is a complicated task. However, physical considerations allow to foresee that $H$ should be comprised between 1-2 meV, a very small value, which in practice prevents its computation by ab-initio methods, as it falls within the accuracy of the highest available levels of theory. The coupling is expected to be small, because it occurs between two heterogeneous (i.e. corresponding to different $\Sigma$ and $\Pi$ symmetries) surfaces, which also ensures the validity of the Landau-Zener approach.
Furthermore, $H$ is the result of two contributions: the spin-orbit coupling, estimated to be slightly larger than in the O+Ar case \cite{Ref4} where the electrostatic contribution is absent, for which the first order non-adiabatic correction to adiabatic potential is around 0.2 meV, and the Coriolis coupling, which should be $\approx$ 1 meV, the value corresponding to a collision with an impact parameter of 1 \AA\ at a relative velocity of 1.5 km/s   \cite{aquilantiCP94}. A value of $H$= 1.5 meV was thus considered in the present calculation.

The radial velocity $v_{R}$ at the crossing in eq. \ref{prob} is given by:
\begin{equation}
v_{R}^{2}=\frac{2}{\mu}\left(E-\frac{\hbar^{2}(l+1) l}{2 \mu R_{c}^{2}}-E_{x}\right),
\end{equation}
in which $l$ is the quantum number representing the orbital angular momentum of the collision complex (from 0 to $l_{\max}$, which guarantees $v_{R}$ to be real), $\mu$ is the reduced mass and $E$ is the collision energy (from 0 to 10 eV). 

The cross section can then be computed from the $P_x$ probability as
\begin{equation}
\sigma(E)=\frac{\pi}{k^{2}} \sum_{l=0}^{l_{\max }}(2 l+1) \cdot 2\left(1-P_{x}\right) P_{x},
\end{equation}
in which $P_{x}$ is the probability of the system staying on the same surface, and (1-$P_{x}$) is that of changing surface. Moreover, $k^{2}=\frac{2 \mu E}{\hbar^{2}}$. The rate coefficient for vibro-electronic energy transfer is obtained by
\begin{small}
\begin{equation}
k_{V-E}(T)=\sqrt{\frac{8 k_{B} T}{\pi \mu}} \frac{1}{\left(k_{B} T\right)^{2}} \int_{0}^{\infty} \sigma(E) e^{-E / k_{B} T} E d E.
\end{equation}
\end{small}

\subsection{The Improved Lennard-Jones model}

For the analytical representation of the $^{3} \Pi$ and $^{3} \Sigma$ PESs, the sum of the first two components, $\left(\mathrm{V}_{v d W}+\mathrm{V}_{c t}\right)$, has been formulated as the combination of pair interactions between $\mathrm{O}\left({ }^{3} \mathrm{P}\right)$ and each N atom of the $\mathrm{N}_{2}$  molecule, represented by an Improved Lennard Jones (ILJ) function.

Specifically, $\left(\mathrm{V}_{v d W}+\mathrm{V}_{c t}\right)$ is described as a sum of atom-“effective atom” contributions involving interaction pair-potentials between $\mathrm{O}\left({ }^{3} \mathrm{P}\right)$ and each N atom of the $\mathrm{N}_{2}$  molecule, i.e.,
\begin{equation}
\label{eq:0}
V_{v d W}+V_{c t}=\sum_{i=1}^{2} V_{N_{i}-O}.
\end{equation}

The contributions in the above equation depend on the “effective” electronic polarizability of the N atom within the $\mathrm{N}_{2}$ molecule, which is different from that of the isolated N atom, and are described by an Improved Lennard Jones (ILJ) function~\cite{pirani2008beyond}, which depends on the distance $R$ between the two interacting centers according to the expression:

\begin{equation}
\begin{aligned}
V_{\mathrm{ILJ}}(R)=& \varepsilon\left[\frac{6}{n(R)-6}\left(\frac{R_{m}}{R}\right)^{n(R)}-\frac{n(R)}{n(R)-6}\left(\frac{R_{m}}{R}\right)^{6}\right],
\end{aligned}
\end{equation}
where $\varepsilon$ and $R_{m}$ (the related parameters are reported in Table S1) are the atom-effective atom interaction well depth and its location, respectively. This function gives a more realistic representation of both the repulsion and the long range attraction than the classic Lennard-Jones potential. The $n$ term is expressed as a function of $R$:
\begin{equation}
n(R)=\beta+4.0\left(\frac{R}{R_{m}}\right)^{2},
\end{equation}
where $\beta$ is a parameter which depends on the hardness of the interacting centers, and it is fixed to 8 in present cases. 

Note that the differences  in $\varepsilon$ and $R_{m}$ potential parameters associated to the two different symmetries, obtained following the guidelines reported in ref. \cite{Ref4}, account  for the electronic anisotropy of the O($^3$P) atom in determining the bond stabilization by CT exclusively in the configuration $^3\Pi$.

\subsection{The quantum-classical method}

The quantum-classical method for atom-diatom collisions was introduced and developed by G.D. Billing \cite{billing1984rate} and is proven to be accurate and efficient to obtain cross sections and rate coefficients of heavy-impact processes involving vibrational energy transfer. 
The key feature of this method is that the vibrational degrees of freedom are treated quantum mechanically, whereas the other degrees of freedom (the translational and the rotational motion) are treated classically.
In order to handle with the whole system in a self-consistent way, the quantum mechanical degrees of freedom must evolve correctly under the influence of the surrounding classical motions. In turn, the classical degrees of freedom must respond correctly to quantum transitions. 

According to the spirit of the quantum-classical method, vibration and rotational-vibrational coupling are treated quantum mechanically by close coupled equations.
For atom-diatom collisions, there is just one quantum degree of freedom (the vibration of the diatom) and the total wavefunction is expanded in terms of the rotationally distorted Morse wave function $\phi_{v}\left(r, t\right)$ as follows:
\begin{equation}
\label{eq:22}
\Psi\left(r, t\right)=\sum_{v} a_{v}(t) \phi_{v}\left(r, t\right) \exp \left(-\frac{i t E_{v}}{\hbar}\right),
\end{equation}
where $r$ is the intramolecular distance of diatom, $E_{v}$ is the eigenvalue of the rotationally distorted Morse wave functions $\phi_{v}\left(r, t\right)$ perturbed by rotational-vibrational coupling 
\begin{equation}
\phi_{v}\left(r, t\right)=\phi_{v}^{0}\left(r\right)+\sum_{v^{\prime} \neq v} \phi_{v^{\prime}}^{0}\left(r\right) \frac{H_{v^{\prime} v}}{E_{v}^{0}-E_{v^{\prime}}^{0}},
\end{equation}
where $H_{v^{\prime} v}$ is the first-order centrifugal stretching term:
\begin{equation}
H_{v^{\prime} v}=-j^{2} m^{-1}\bar{r}^{-3}<\phi_{v^{\prime}}^{0}| r-\bar{r} | \phi_{v}^{0}>
\end{equation}
with $j$ being the rotational momentum of the molecule and the operator $<>$ is obtained by integrating over $r$. $\phi_{v}^{0}$ is the unperturbed eigenfunction of the Morse oscillator and $E_{v}^{0}$ is the eigenvalue approximated as
\begin{equation}
E_{v}^{0}=\hbar \omega_{e}\left(v+\frac{1}{2}\right)-\hbar \omega_{e} x_{e}\left(v+\frac{1}{2}\right)^{2}+ \hbar \omega_{e} y_{e}\left(v+\frac{1}{2}\right)^{3},
\end{equation}
where $\omega_{e}$ is the oscillator wavenumber and $x_{e}$ and $y_{e}$ are the anharmonicity constants. 

In order to obtain the amplitudes $a_{v^{\prime} }$ for the inelastic processes N$_{2}\left(v\right)+$O$ \rightarrow$ N$_{2}\left(v^{\prime}\right)+$O, one then plugs the expansion (eq. \ref{eq:22}) into the time-dependent Schr\"odinger equation and has to solve the following set of coupled equations for the amplitudes:
\begin{equation}
\label{eq:3}
\begin{aligned} i \hbar \dot{a}_{v^{\prime} }(t) &=\sum_{v}\left[\left\langle\phi_{v^{\prime}}^{0} \left|V\left(R,r, \gamma\right)+ 2 i \hbar j \frac{d j}{d t} \frac{\left\langle\phi_{v^{\prime}}^{0}\left|\left(r-\bar{r}\right)\right| \phi_{v}^{0}\right\rangle}{m r_{e q}^{3}\left(E_{v}^{0}-E_{v^{\prime}}^{0}\right)} \right| \phi_{v}^{0} \right\rangle\right]  \cdot a_{v}(t) \exp \left[\frac{i}{\hbar}\left(E_{v^{\prime}}-E_{v}\right) t\right], \end{aligned}
\end{equation}
in which the intermolecular potential $V(R, r, \gamma)$ is conveniently expressed as a function of $R$, the distance between the atom and the center of mass of the diatom and $\gamma$ (the angle between $r$ and $R$). The translational and rotational motions are obtained by solving the corresponding Hamilton equations by making use of an Ehrenfest averaged potential~\cite{billing1984semiclassical} defined as the quantum expectation value of the interaction potential. This mean-field method usually provides accurate quantum transition probabilities and properly conserves total (quantum plus classical) energy. 
A variable-order variable-step Adams predictor-corrector integrator~\cite{hamming1986numerical} is then used to solve the coupled equations (eq. \ref{eq:3}) and the classical equations of motion for rotation and translation. An absolute integration accuracy of $10^{-8}$ is achieved for all calculations in this work.

The vibrational wavefunction is initialized as a Morse wavefunction. The simultaneous propagation of the quantum and classical sets of equations produces the quantum transition amplitudes $a_{v^{\prime}}$ which can be used to calculate cross sections for the vibrational transitions.  The cross sections are obtained by averaging over a number of trajectories having randomly selected initial conditions, and a Monte Carlo average over the initial Boltzmann distribution of rotational energy is introduced to have rate coefficients for vibrational energy transfer. Thus an averaged cross section is defined as:
\begin{equation}
\begin{aligned} \sigma_{v \rightarrow v^{\prime} }\left(T_{0}, \bar{U}\right)=& \frac{\pi \hbar^{4}}{4 \mu k_{B}^{2} T_{0}^{2} I} \int_{0}^{J_{\max }} \int_{0}^{j_{m a x}} \int_{l=|J-j|}^{J+j} d J d j d l \cdot\left(2 J+1\right) P_{v \rightarrow v^{\prime} }, \end{aligned}
\end{equation}
where $\mu$ is the reduced mass for the relative motion, $J$ the total, $j$ the rotational and $l$ the initial orbital angular momentum. The moment of inertia is $I=m r^{2}$ and the temperature $T_{0}$ is arbitrary because it cancels out when calculating the rate coefficients. $J_{m a x}$ and $j_{m a x}$ are the upper limit for the randomly chosen total and rotational quantum numbers. Rate coefficients are then calculated through the following equation
\begin{equation}
\begin{aligned} k_{v \rightarrow v^{\prime}}(T) &=\left(\frac{8 k_{B} T}{\pi \mu}\right)^{1 / 2}\left(\frac{T_{0}}{T}\right)^{2} \int_{0}^{\infty} d\left(\frac{\bar{U}}{k_{B} T}\right) \cdot\exp \left(-\frac{\bar{U}}{k_{B} T}\right) \sigma_{v \rightarrow v^{\prime} }\left(T_{0}, \bar{U}\right), \end{aligned}
\end{equation}
which holds for exothermic processes. $\bar{U}$, the symmetrized classical energy, is introduced to restore the detailed balance principle.\cite{billing1984rate,billing1984semiclassical}

Molecular parameters used in the present calculations are reported in Table S2.

\newpage
\noindent

\begin{table}
{\bf Table S1:}
 Parameters of the analytical formulation of the $^3\Pi$  and $^3\Sigma$ PESs for N$_2$ at its equilibrium distance (r$_e$=1.1007 \AA). Note that parameters involved in the ILJ functions refer to atom - effective atom additive components.
 \label{tab:S1}
 \vspace{0.4cm}
\setlength{\tabcolsep}{12pt}
\centering
 \begin{tabular}{l|c|c}
%  &\multicolumn{6}{c}{R$\rightarrow$TS}  \\
 \hline
  &  $^3\Pi$ & $^3\Sigma$ \\
 \hline
  &\multicolumn{2}{c}{O-N interaction (ILJ formulation\cite{Ref3})}  \\
 \hline
 $\varepsilon$ (meV)  & 7.70 & 3.10    \\
 $R_m$  (\AA) &  3.32  & 3.87    \\ 
 $\beta$        &  8   &  8  \\
\\
  &\multicolumn{2}{c}{ quadrupole-quadrupole interaction}  \\
 \hline
 Q$_O$ (a.u)  &  0.475    &  -0.950 \\
 Q$_{N_2}$ (a.u.) &  -1.115  &  -1.115   \\
\hline
\end{tabular}

\vspace{2cm}
\end{table}

\begin{table}[h]

{\bf Table S2:} Molecular constants for N$_{2}$.
\vspace{0.4cm}
\centering
  \label{tal:table1}
  
  \begin{tabular*}{0.48\textwidth}{@{\extracolsep{\fill}}ll}
    \hline
$\omega_{e}$ &  2359.60  cm$^{-1}$ \\
$x_{e}$ &  0.006126 \\
$y_{e}$ &  0.0000032 \\
$r_{e q}$ &  1.1007 \AA \\
$\beta$ &  2.689 \AA$^{-1}$ \\
$D_e$ & 9.905 eV \\
    \hline
  \end{tabular*}
\end{table}

\vspace{2cm}

\begin{table*}[h]
\centering
{{\bf Table S3}: Rate coefficients for total vibrational relaxation of $\mathrm{N}_{2}\left({ }^{1} \Sigma_{g}^{+}\right)(v=1)$ upon collision with $\mathrm{O}\left({ }^{3} P\right)$ as a function of temperature.}
\vspace{0.4cm}

%\begin{small}
%\begin{tabular}{ccc}
 \begin{tabular*}{0.48\textwidth}{@{\extracolsep{\fill}}ccc}
   \hline
 T & V-T+V-E rates & Expt. \cite{mcneal} \\ \hline
 300 & 1.05E-15 & 3.32E-15 \\
 460 & 9.95E-15 & 1.39E-14 \\
 640 & 3.22E-14 & 2.91E-14 \\
 740 & 4.83E-14 & 4.43E-14 \\
    \hline
\end{tabular*}
%\end{small}
\end{table*}
\newpage

\begin{figure}
\centering
\includegraphics[width=11.cm,angle=0.]{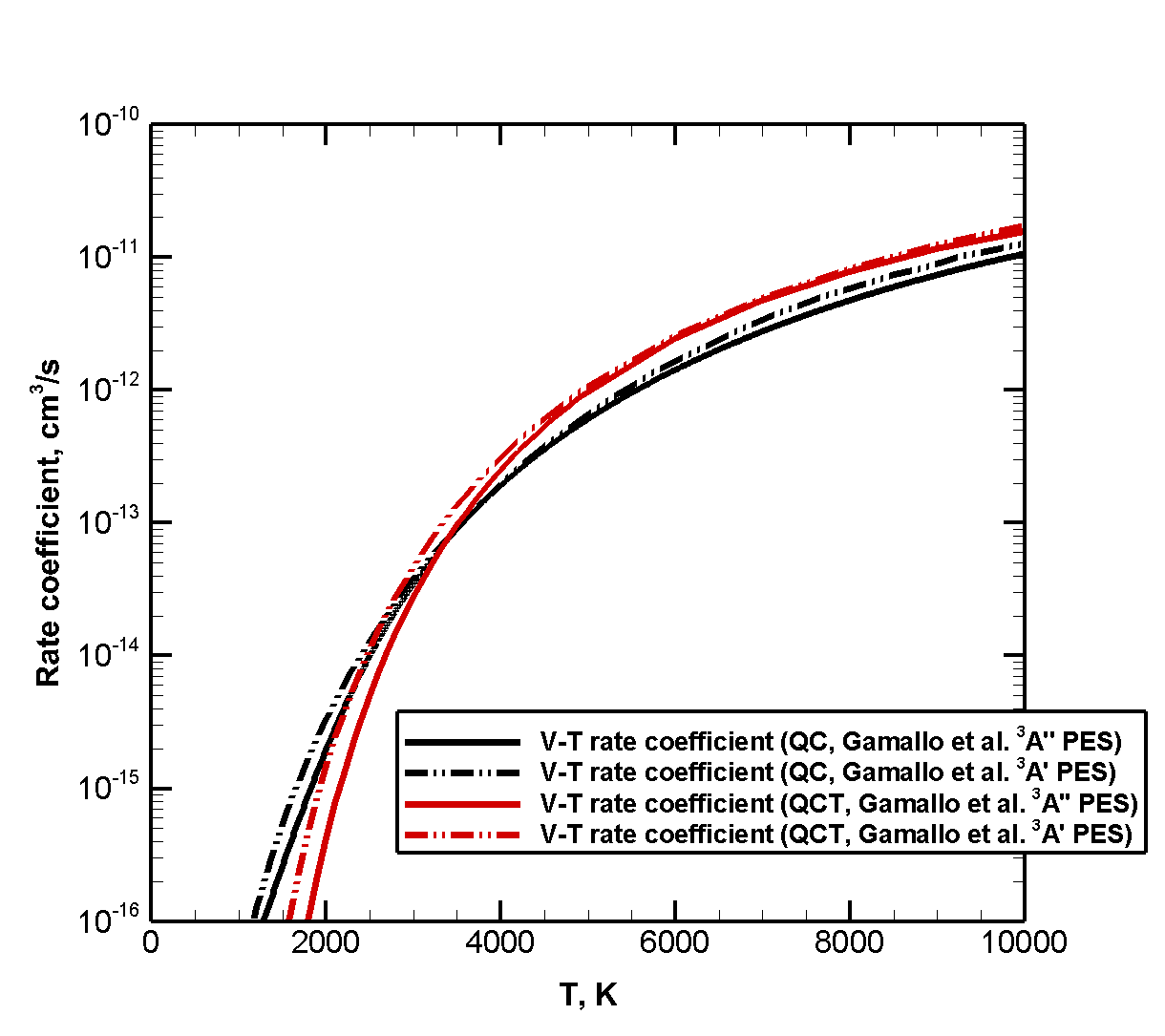}\\
{\bf Figure S1}: Rate coefficients for vibrational relaxation of  N$_2$($^1\Sigma_g^+$)($v=1$) upon collision with O($^3P$) as a function of temperature. V-T rate coefficients computed on Gamallo et al. \cite{sayos} $^3$A$^{''}$  (solid lines) and  $^3$A$^{'}$ PESs (dashed lines) by QCT \cite{esposito2} (red) and by the present QC (black) methods.
%\label{fig2}
\end{figure}

\begin{figure}
\centering
\includegraphics[width=11.cm,angle=0.]{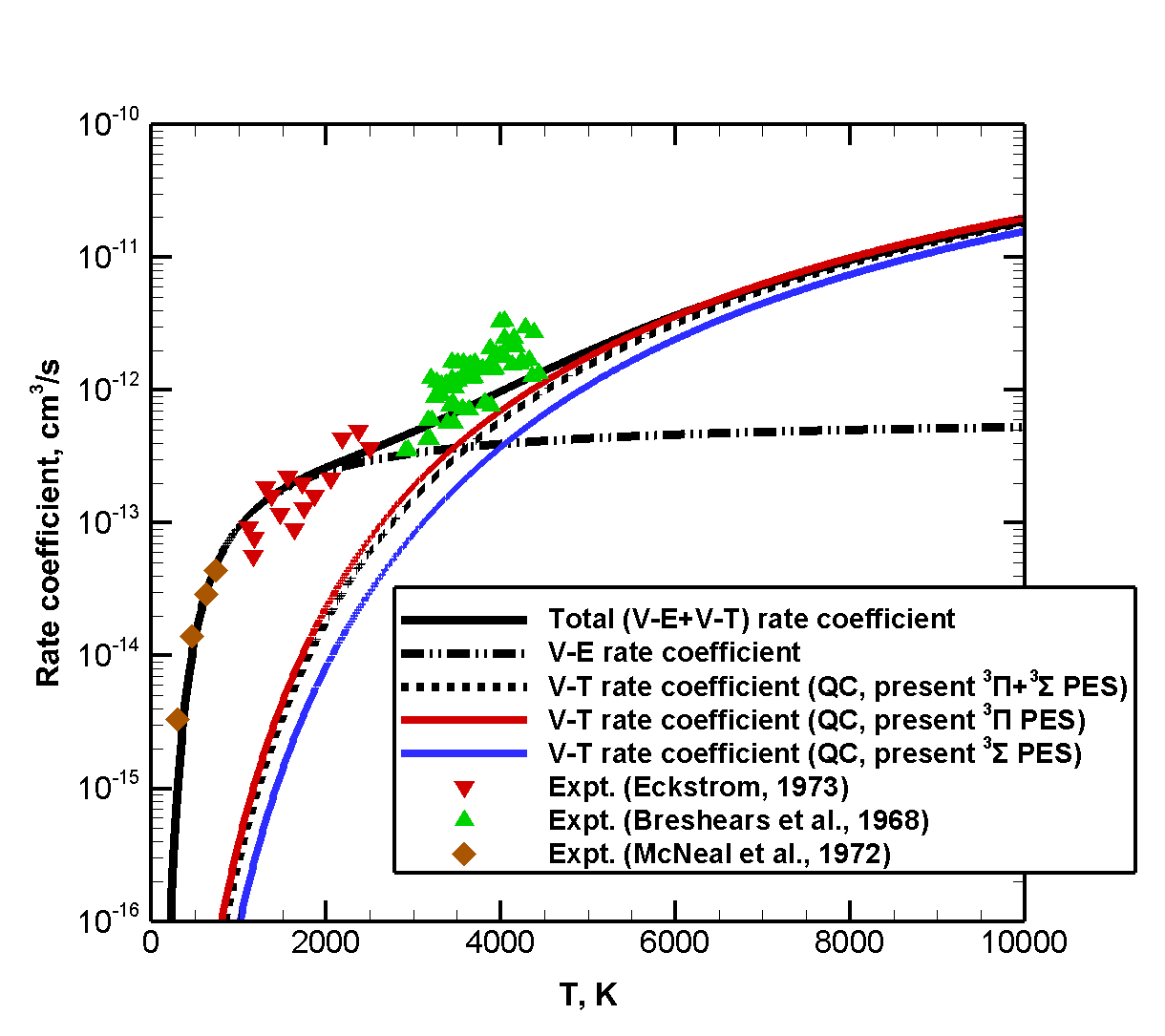}\\

%\label{fig1}

{\bf Figure S2}: Rate coefficients for vibrational relaxation N$_2$($^1\Sigma_g^+$)($v=1$) upon collision with O($^3P$) as a function of temperature. Experimental data  by Eckstrom \cite{eckstrom} (red down triangles), by Breshears et al. \cite{breshears} (green up triangles) and by McNeal \cite{mcneal} (brown diamonds) are reported together with QC V-T rate coefficients computed on the present $^3\Pi$  (red solid line) and  $^3\Sigma$ (blue solid line): averaged (2:1) V-T rate coefficients are also reported (black dashed  line), together with V-E ( black dash-dot line) and total (V-T+V-E) vibrational relaxation rate coefficients (black solid line).
\end{figure}

\begin{figure}
\centering
\includegraphics[width=15.cm,angle=0.]{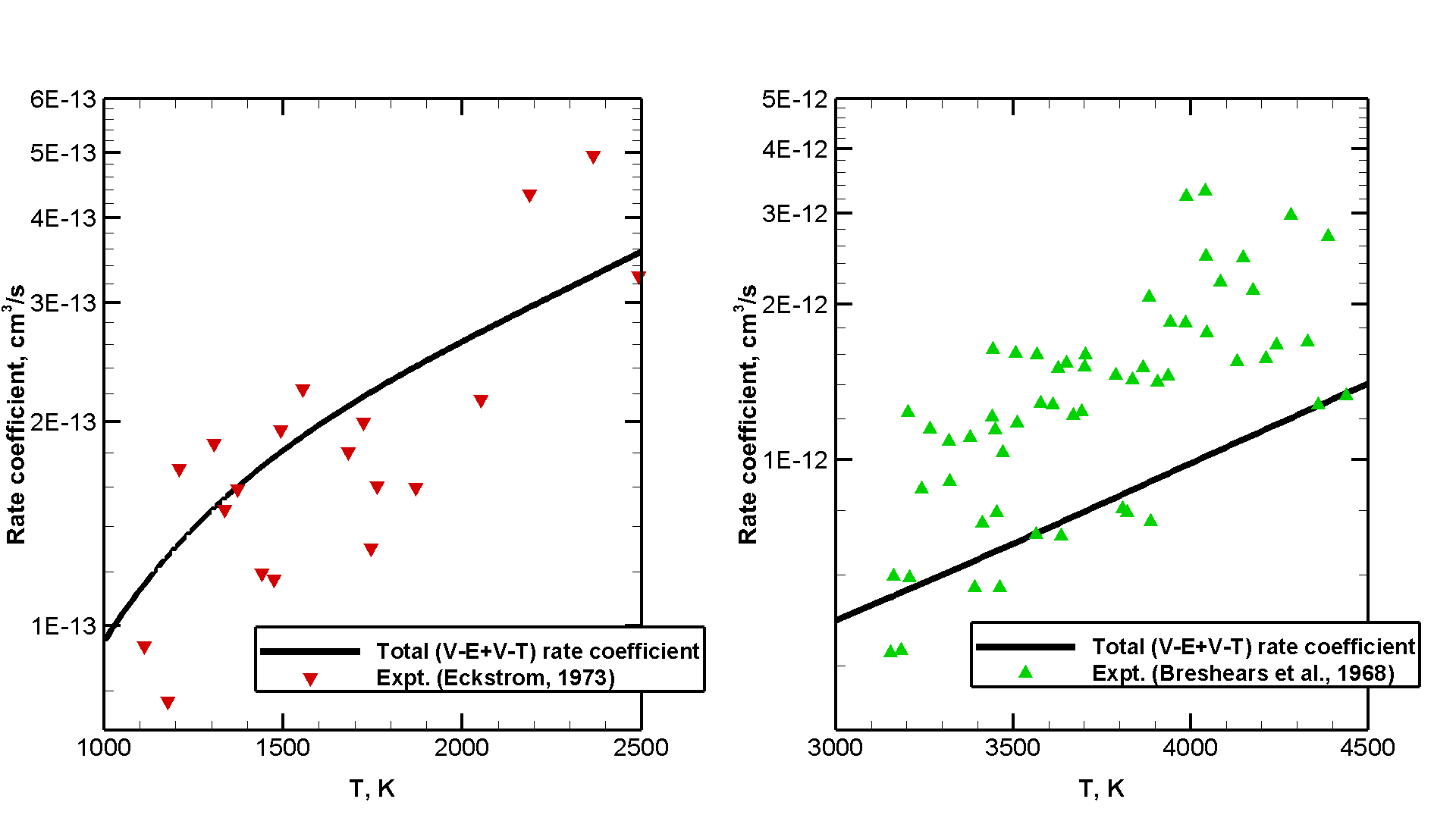}\\

%\label{fig3}

{\bf Figure S3}: Calculated rate coefficients for vibrational relaxation N$_2$($^1\Sigma_g^+$)($v=1$) upon collision with O($^3P$) as a function of temperature in the 1000-2500 K range, left panel, and 3000-4500 K range, right panel. Experimental data  by Eckstrom \cite{eckstrom} (red down triangles) and by Breshears et al. \cite{breshears} (green up triangles)  are reported.
\end{figure}

\begin{figure}
\centering
\includegraphics[width=11.cm,angle=0.]{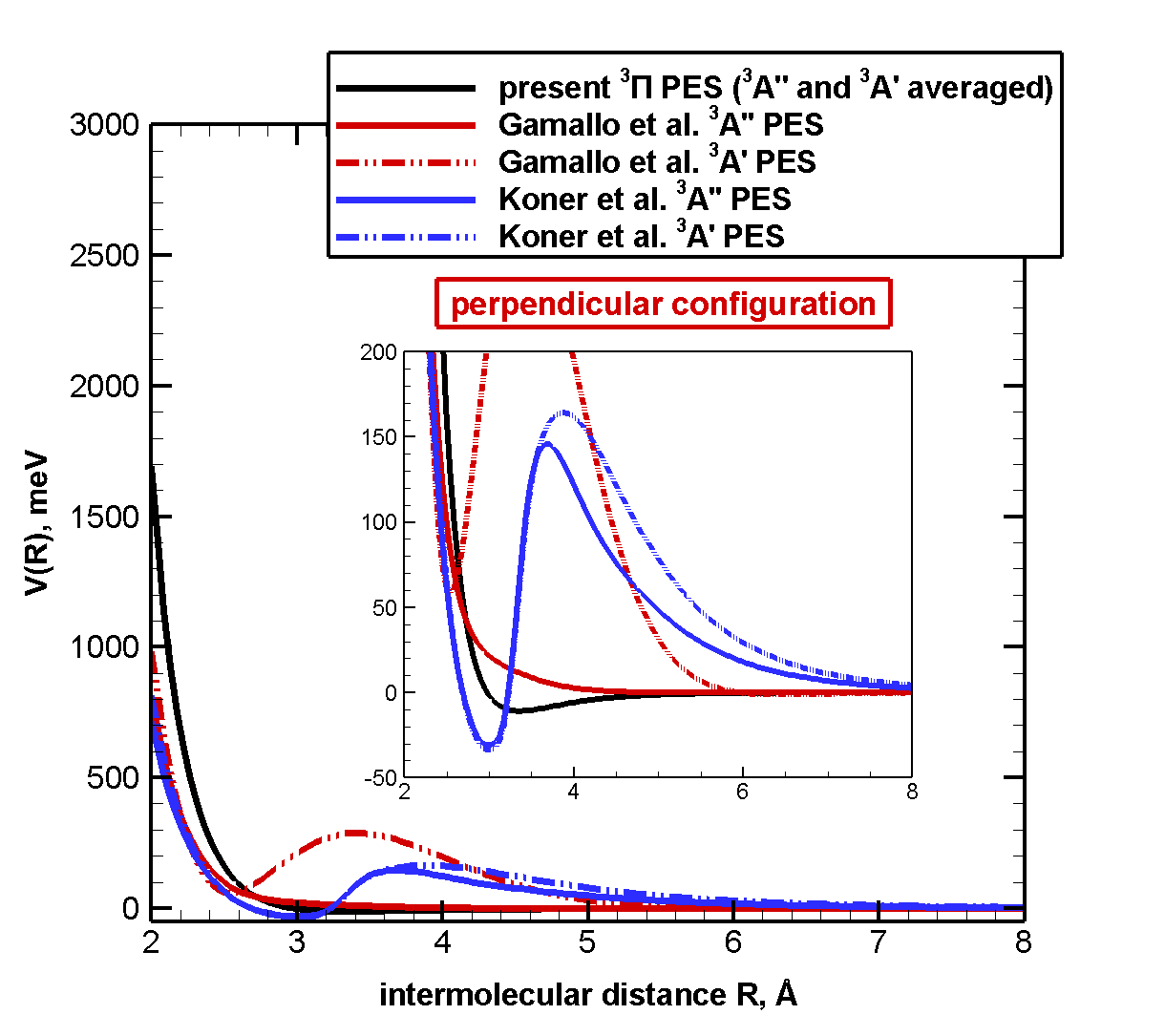}\\
{\bf Figure S4}: Behavior of different potential energy surfaces as a function of the intermolecular distance $R$ for the perpendicular configuration, corresponding to the  C$_{2v}$ symmetry. The present $^3\Pi$ PES is reported as a solid black line, the Gamallo et al. \cite{sayos} $^3$A$^{''}$ and $^3$A$^{'}$ are the red solid and dashed lines, respectively, and the Koner et al. \cite{meuwly2020b} $^3$A$^{''}$ and $^3$A$^{'}$ are the blue solid and dashed lines, respectively.
%\label{fig4}
\end{figure}  

\clearpage

%\bibliographystyle{unsrt}  
%\bibliography{apssamp}  %%% Remove comment to use the external .bib file (using bibtex).
%%% and comment out the ``thebibliography'' section.

\providecommand{\noopsort}[1]{}\providecommand{\singleletter}[1]{#1}%

\end{document}